\documentclass[12pt]{iopart}
\usepackage{epsfig}
\eqnobysec
\newcommand{\curl}{\,\mbox{curl}\,}

\newcommand{\sigt}{\sigma_{\mathrm{T}}}
\newcommand{\nelec}{n_{\mathrm{e}}}
\newcommand{\ud}{{\mathrm{d}}}
\newcommand{\uD}{{\mathrm{D}}}
\newcommand{\tc}{t_{\mathrm{c}}}
\newcommand{\clq}{{\mathcal{Q}}}
\newcommand{\TT}{{\mathrm{TT}}}
\newcommand{\clh}{{\mathcal{H}}}
\newcommand{\clp}{{\mathcal{P}}}
\begin{document}
\jl{6}
\title[Microwave background anisotropies from gravitational waves]{Microwave
background anisotropies from gravitational waves: the 1+3 covariant approach}
\author{Anthony Challinor\footnote{A.D.Challinor@mrao.cam.ac.uk}}
\address{Astrophysics Group, Cavendish Laboratory, Madingley Road,
Cambridge CB3 0HE, UK}
\begin{abstract}
We present a 1+3 covariant discussion of the contribution of gravitational
waves to the anisotropy of the cosmic microwave background radiation (CMB) in
an almost-Friedmann-Robertson-Walker (FRW) universe. Our discussion
is based in the covariant approach to perturbations in cosmology, which
provides a physically transparent and gauge-invariant methodology
for CMB physics. Applying this approach to linearised gravitational waves,
we derive a closed set of covariant equations describing the evolution
of the shear and the Weyl tensor, and the angular multipoles of the CMB
intensity, valid for an arbitrary matter description and
background spatial curvature. A significant feature of the present approach
is that the normal mode expansion of the radiation distribution function,
which arises naturally here, preserves the simple quadrupolar nature of the
anisotropic part of the Thomson scattering source terms, and provides a direct
characterisation of the power in the CMB at a given angular multipole, as in
the recently introduced total angular momentum method. We provide the integral
solution of the multipole equations, and analytic solutions for the shear
and the Weyl tensor, for models with arbitrary spatial curvature. Numerical
results for the CMB power spectrum in open models are also presented,
which provide an independent verification of the calculations of other groups.
\end{abstract}
%
\section{Introduction}

The cosmic microwave background radiation (CMB) plays an essential
role in modern cosmology. Improved observations of the structure in the CMB
should allow one to distinguish between rival theories of structure formation,
and, in the case of the inflationary scenario, to determine the spectrum of
primordial scalar (density) and tensor (gravitational wave) perturbations.
Vector perturbations are expected to leave a significant
imprint on the CMB only in the presence of active seeds, such as the defect
models considered in~\cite{pen97}.

The calculation of CMB anisotropies in specific cosmological models
is now a large industry. For a representative sample of the literature
dealing with the imprint of scalar perturbations, see
\nocite{sachs67,peebles70,wilson83,bond84,abbott86,ma95,seljak96}
e.g.~\cite{sachs67}--\cite{seljak96}; for tensor perturbations see
\nocite{dautcourt69,rubakov82,fabbri83,abbott84,white92,critt93b,atrio94,frewin94,allen95,hu97b,durrer98}
e.g.~\cite{dautcourt69}--\cite{durrer98}.
Early calculational schemes were based on
linear perturbation theory in the synchronous gauge~\cite{lifshitz46}, with
Legendre expansions of the relativistic distribution
functions~\cite{peebles70,bond84}.
Some technical improvements in these schemes resulted from the use of
gauge-invariant variables~\cite{abbott86,bardeen80}, and, more recently,
from improved normal mode expansions of the distribution
functions~\cite{hu97a,hu98}.
Adopting gauge-invariant variables frees one from the need to keep track of
the implications of any residual gauge freedom in the calculation (the
so called gauge problem; see e.g.~\cite{ellis89a,bruni99}),
while improvements
in the normal mode expansions have led to more direct characterisations of the
anisotropy in vector and tensor modes, and a simpler form for the anisotropic
Thomson scattering terms than is possible with the traditional
Legendre expansion.

Recently, the 1+3 covariant and gauge-invariant approach to perturbations in
cosmology, developed by Ellis and coworkers~\cite{ellis89a,ellis89b} from
the covariant hydrodynamics of Ehlers~\cite{ehlers93}, has been applied to
problems in CMB
\nocite{maartens95,stoeger95b,dunsby96b,LC-sw,LC-scalcmb,gebbie99a,maartens99,
chall99a,gebbie99b}
physics~\cite{maartens95}--\cite{gebbie99b}. The 1+3 covariant approach
provides a very
convenient framework within which to study the exact intrinsic
dynamics of cosmological models. The approach is formulated at the level of
variables that are covariantly defined in the real universe, and are
(exactly) gauge-invariant by construction, thus freeing one from many of the
limitations of previous methodologies. Furthermore, the approach admits a
covariant and gauge-invariant linearisation that allows linearised
calculations to be performed in a very direct manner. In~\cite{LC-scalcmb}
a general 1+3 covariant scheme for the calculation of linearised CMB
anisotropies was developed (see also~\cite{gebbie99a,gebbie99b}).
A complete, closed set of equations, that fully
describe the linearised evolution of inhomogeneity and anisotropy in an
almost-Friedmann-Robertson-Walker (FRW) universe with arbitrary spatial
curvature, were presented. These equations
were subsequently extended to include non-linear terms in~\cite{maartens99},
where some qualitative implications of non-linear effects were also noted.
A significant feature of the covariant approach is the unified way in
which scalar, vector and tensor perturbations are handled; the
general perturbation equations are formulated without recourse to a
decomposition into scalar, vector and tensor modes, although such a
decomposition can still be made, if required, at a late stage in the
calculation.
In~\cite{LC-scalcmb} the detailed equations describing
scalar modes were derived from the unified equations, along with the
necessary machinery to calculate the power spectrum of the CMB temperature
anisotropy from a solution of these equations. It is our purpose here to
report the extension of the 1+3 covariant analysis to tensor modes. We shall
see that, in addition to the advantages of gauge-invariance and physical
clarity that are evident in the scalar calculation, the
tensor calculation highlights further advantages of the
covariant approach over more traditional treatments. Namely, in the present
approach a normal mode expansion of the radiation distribution function arises
naturally that has the virtue of maintaining a direct characterisation
of the power in the CMB at a given multipole $\ell$, and of preserving the
simple quadrupolar nature of that part of the Thomson scattering source terms
that arises from the angular dependence of the Thomson cross section. Similar
advantages are shared by the recently introduced total angular momentum
method~\cite{hu97a,hu98}.

The paper is arranged as follows. In section~\ref{sec_cov} we briefly review
the 1+3 covariant approach to cosmological perturbations in an almost-FRW
model, radiative transfer, and the decomposition of the CMB sky in
angular multipoles. Section~\ref{sec_gw}
begins with a discussion of the covariant description of
gravitational waves in cosmology, and presents linearised wave equations,
valid for a general matter stress-energy tensor, for the electric and magnetic
parts of the Weyl tensor and the shear. These equations extend the
perfect-fluid equations given in~\cite{dunsby96a}. Following an expansion
in tensor harmonics, in section~\ref{sec_sol} we give analytic solutions for
the shear and the Weyl tensor during matter and radiation domination, extending
the solutions given in~\cite{dunsby96a} to the case of non-flat spatial
sections. We also provide the integral solutions to the mode-expanded
radiation multipole equations in the general case.
In section~\ref{sec_pow} we derive the CMB power spectrum in terms of the
mode-expanded multipole moments, and present numerical solutions in open CDM
models, verifying the results in~\cite{hu97b}.
We close in section~\ref{sec_conc} with our conclusions.
\ref{appa} contains useful new results on the symmetric trace-free
representation of the tensor harmonics which we use in
sections~\ref{sec_sol} and~\ref{sec_pow}.

We employ standard general relativity with a $(+---)$ metric signature.
Our conventions for the Riemann and Ricci tensors are fixed by
$2\nabla_{[a},\nabla_{b]} u_c = R_{abc}{}^d u_d$, and
$R_{ab} \equiv{R_{acb}}^c$, where $\nabla_a$ denotes the covariant derivative.
Round brackets denote symmetrisation on the enclosed indices, square brackets
denote antisymmetrisation, and angle brackets denote the projected symmetric
trace-free (PSTF) part. We use units with $c=G=1$ throughout, so that the
constant in the Einstein equation is $\kappa = 8\pi$. It should be noted that
our convention for the metric signature, which coincides with that in our
earlier work on CMB anisotropies in the 1+3 covariant
approach~\cite{LC-sw,LC-scalcmb,chall99a}, differs from much of the 1+3
covariant literature. It is straightforward to transform from our
conventions to those in~\cite{maartens99}: $g_{ab} \rightarrow -g_{ab}$,
$\nabla_a \rightarrow \nabla_a$, and so $R_{abc}{}^d \rightarrow
R_{abc}{}^d$. It follows that $R_{ab}\rightarrow R_{ab}$, but for the
Ricci scalar $R\rightarrow -R$. The matter stress-energy tensor $T_{ab}$
is unchanged, as is the fundamental velocity (vector) field $u^a$ of the
cosmological model. Raising or lowering an index incurs a sign change when
transforming between signatures. The transformations of other key geometric
objects are discussed in the text as they are introduced.

\section{The 1+3 covariant approach to perturbations in cosmology}
\label{sec_cov}

The essence of the 1+3 covariant approach
is to identify a set of covariant variables which conveniently describe the
inhomogeneity and anisotropy of the universe in a physically transparent
(gauge-invariant) manner. Having chosen such variables, exact equations
governing their dynamics can be found from the full field equations. Since the
variables are chosen to have a direct physical interpretation, the physical
content of their dynamical equations is often simple to extract. If required,
the exact equations can be linearised around a chosen background to yield
a simple set of perturbation equations that describe the evolution of
physically relevant measures of the inhomogeneity and anisotropy of the
universe in a very direct fashion.

The approach employs a 1+3 decomposition of geometric quantities with respect
to a (fundamental) timelike velocity field $u^a$. We require that, in a general
cosmological model, $u^a$ be chosen in a physical manner such that in the
FRW limit it correctly reduces to the preferred velocity. This restriction
is necessary to ensure gauge-invariance of the approach. Tensors are
decomposed irreducibly with respect to $u^a$ into projected tensors
which are orthogonal to $u^a$ on every index. For example, the
matter stress-energy tensor $T_{ab}=T_{(ab)}$ decomposes
as\footnote{To transform to the conventions of~\cite{maartens99}, one should
make the replacements $q_a \rightarrow -q_a$, $\pi_{ab} \rightarrow
\pi_{ab}$, as well as $u_a \rightarrow -u_a$ and $h_{ab} \rightarrow -h_{ab}$.
Of course, $\rho$ and $p$ are unchanged.}
\begin{equation}
T_{ab} = \rho u_a u_b + 2 u_{(a}q_{b)} - p h_{ab} + \pi_{ab},
\end{equation}
which defines the energy density $\rho \equiv T_{ab} u^a u^b$, the
energy flux $q_a \equiv h^b_a T_{bc} u^c$, the isotropic pressure
$p\equiv - h^{ab} T_{ab} /3$, and the anisotropic stress $\pi_{ab}
\equiv T_{\langle ab \rangle}$ which is the projected symmetric trace-free
(PSTF) part of $T_{ab}$. Here, $h_{ab} \equiv g_{ab} - u_a u_b$ is the
projection tensor into the instantaneous rest space of $u^a$ (and $g_{ab}$ is
the metric of spacetime). In the FRW limit $q_a$ and $\pi_{ab}$ vanish
since the stress-energy tensor is then restricted to perfect-fluid form.

It is convenient to introduce a projected covariant ``derivative''
$\uD_a$ which acts on a general tensor ${T_{b\dots}}^{c\dots}$ according to
\begin{equation}
\uD_a {T_{b\dots}}^{c\dots} \equiv h^d_a h^e_b \dots h^c_f \dots
\nabla_d {T_{e\dots}}^{f\dots}.
\end{equation}
It should be noted that $\uD_a$ is a derivation only when acting on projected
objects. When $u^a$ is irrotational (see below), $\uD_a$ reduces to the
covariant derivative in the hypersurfaces orthogonal to $u^a$.
It is also convenient to introduce the derivative along the flow lines
of $u^a$, denoted by an overdot, so that ${\dot{T}_{b\dots}}^{\phantom{b\dots}
c\dots} = u^a \nabla_a {T_{b\dots}}^{c\dots}$.
With these definitions, the covariant derivative of $u^a$ decomposes
as\footnote{To transform to the metric signature of~\cite{maartens99}, make the
replacements $A_a \rightarrow -A_a$, $\sigma_{ab} \rightarrow -\sigma_{ab}$,
$\omega_{ab} \rightarrow \omega_{ab}$, as well as $u_a \rightarrow -u_a$,
$\nabla_a \rightarrow \nabla_a$ and $h_{ab} \rightarrow -h_{ab}$.
It follows that for the vorticity
vector $\omega_a \rightarrow \omega_a$. Our $\omega_a$
is physically minus that in~\cite{maartens99} (rather than just any
difference due to the metric signature), but coincides physically
with that in~\cite{ellis98}.}
\begin{equation}
\nabla_a u_b=u_a A_b +\sigma_{ab} + {\frac{1}{3}}\Theta h_{ab} + \omega_{ab},
\end{equation}
which defines the acceleration $A_a \equiv \dot{u}_a$, the
shear $\sigma_{ab} \equiv\uD_{\langle a} u_{b \rangle}$, the (volume) expansion
scalar $\Theta = \uD^a u_a = 3H$, with $H$ the local Hubble parameter, and the
vorticity $\omega_{ab} \equiv \uD_{[a}u_{b]}$.
Introducing a  projected, totally antisymmetric tensor $\epsilon_{abc}
\equiv \eta_{abcd}u^d$, where $\eta_{abcd}$ is the totally antisymmetric
tensor on spacetime, we can define the vorticity (projected) vector
as the dual of $\omega_{ab}$: $\omega_a \equiv \epsilon_{abc} \omega^{bc} /2$.
It is also convenient to generalise the notion of the 3-dimensional curl
to tensors ${T_{ab\dots c}}$ via
\begin{equation}
\curl T_{ab\dots c} \equiv \epsilon_{d e (a} \uD^d {T_{b\dots c)}}^e,
\end{equation}
which is PSTF for ${T_{ab\dots c}}$ PSTF. The vorticity vector can then
be written as $\omega_a = \curl u_a /2$. The kinematic variables 
$A_a$, $\omega_a$ and $\sigma_{ab}$ characterise anisotropy in the
local expansion and thus vanish identically in an exact FRW universe. The
projected gradient of the expansion scalar, $\uD_a \Theta$, describes
inhomogeneity in the expansion. Like the projected derivative of any
covariant object, $\uD_a \Theta$ vanishes in the FRW limit.

The locally free part of the gravitational field is described by the
Weyl tensor $C_{abcd}$. The ten degrees of freedom in $C_{abcd}$ can be
encoded in two PSTF tensors, $E_{ab}$ and $H_{ab}$, denoted the electric
and magnetic parts, respectively:\footnote{Replace $E_{ab} \rightarrow
-E_{ab}$ and $H_{ab} \rightarrow -H_{ab}$, and also $C_{acbd}\rightarrow
-C_{acbd}$, $u^a \rightarrow u^a$, and $\epsilon_{acd} \rightarrow
\epsilon_{acd}$ to transform to the conventions of~\cite{maartens99}.}
\begin{eqnarray}
E_{ab}  \equiv  u^c u^d C_{acbd} \\
H_{ab}  \equiv  \frac{1}{2} \epsilon_{acd} {C_{be}}^{cd} u^e.
\end{eqnarray}
The electric part of the Weyl tensor plays an analogous role to the
tidal tensor derived from the potential in Newtonian gravitation, while
$H_{ab}$ is essential for the propagation of gravitational waves. Indeed,
the non-vanishing of the super-energy flux~\cite{maartens98} $P_a \equiv
{\epsilon_{ac}}^d E^{bc} H_{bd}$ for any timelike $u^a$
forms the basis of Bel's first
condition for the existence of gravitational radiation in an (exact)
solution of the field equations (see e.g.~\cite{zak-gw}).
Since the Weyl tensor vanishes in the FRW limit, $E_{ab}$ and $H_{ab}$
provide a covariant and gauge-invariant characterisation of the
perturbation in the free gravitational field.

To construct a covariant linear perturbation theory about an FRW universe,
those covariant variables that vanish in an exact FRW universe are
regarded as being $O(\epsilon)$ in a smallness parameter $\epsilon$.
The linearisation procedure then consists of neglecting products of
$O(\epsilon)$ variables in the dynamical equations. The exact dynamical
equations, which follow from the field equations, are given
in~\cite{maartens99,ellis98}.
Here, we shall only require the equations in
linearised form around an FRW universe. There are seven propagation
equations describing evolution along the flow lines of $u^a$:
\begin{eqnarray}
\fl\dot{\rho} & = & - (\rho + p) \Theta - \uD^a q_a \label{eq:p1} \\
\fl\dot{\Theta} & = & -{\frac{1}{3}}\Theta^2 - {\frac{1}{2}}\kappa
(\rho + 3p) + \uD^a A_a \label{eq:p2} \\
\fl\dot{q}_a & = & -{\frac{4}{3}} \Theta q_a - (\rho + p)A_a + \uD_a p
-\uD^b\pi_{ab}\label{eq:p3} \\
\fl\dot{\omega}_a & = & - {\frac{2}{3}}\Theta \omega_a + {\frac{1}{2}}
\curl A_a \label{eq:p4} \\
\fl\dot{\sigma}_{ab} & = & - {\frac{2}{3}} \Theta \sigma_{ab} - E_{ab}
- {\frac{1}{2}}\kappa \pi_{ab} + \uD_{\langle a}A_{b\rangle} \label{eq:p5}\\
\fl\dot{E}_{ab} & = & -\Theta E_{ab} + \curl H_{ab} + {\frac{1}{2}}
\kappa \left[ -(\rho + p)\sigma_{ab} - \uD_{\langle a} q_{b\rangle}
+ \dot{\pi}_{ab}+ {\frac{1}{3}} \Theta \pi_{ab}\right] \label{eq:p6} \\
\fl\dot{H}_{ab} & = & - \Theta H_{ab} - \curl E_{ab} - {\frac{1}{2}}\kappa
\curl \pi_{ab}, \label{eq:p7}
\end{eqnarray}
and five constraint equations:
\begin{eqnarray}
\uD^a \omega_a  =  0 \label{eq:c1}\\
\uD^a \sigma_{ab} - \curl \omega_b - {\frac{2}{3}}\uD_b \Theta - \kappa q_b
= 0 \label{eq:c2}\\
\uD^a E_{ab} - \kappa \left({\frac{1}{3}}\Theta q_b + {\frac{1}{3}}
\uD_b \rho + {\frac{1}{2}} \uD^a \pi_{ab}\right) =0 \label{eq:c3}\\
\uD^a H_{ab} - {\frac{1}{2}}\kappa [ 2(\rho + p)\omega_b + \curl q_b ]=0
\label{eq:c4} \\
H_{ab} - \curl \sigma_{ab} + \uD_{\langle a} \omega_{b \rangle}=0,
\label{eq:c5}
\end{eqnarray}
which serve to constrain the initial data. It is straightforward to verify
that the constraint equations are preserved by the propagation equations.

We describe the CMB radiation field via the specific intensity $I(E,e^a)$
which gives the intensity per unit solid angle along the (projected) direction
$e^a$ at photon energy $E$. The energy and propagation direction relative to
$u^a$ are related to the four-momentum of the photon by
\begin{equation}
p^a = E(e^a + u^a).
\end{equation}
The photon distribution function $I(E,e^a)/E^3$ is frame-independent for given
$p^a$. It is convenient to work with the energy-integrated intensity,
\begin{equation}
I(e^a) = \int_0^\infty I(E,e^a) \, \ud E,
\end{equation}
since this determines the bolometric temperature along a given direction
$e^a$. Denoting the fractional temperature difference from the all-sky average
at some point by $\delta_T(e^a)$, we have
\begin{equation}
[1+ \delta_T(e^a)]^4 = \left({\frac{1}{4\pi}}\int I(e^a)\,\ud\Omega\right)^{-1}
I(e^a), \label{eq:0}
\end{equation}
where the integration is over solid angles.

The anisotropy of the CMB intensity is conveniently described by
the PSTF multipole moments $I_{A_\ell}$\footnote{We use a lumped index
notation, $A_\ell = a_1 \dots a_\ell$.} in the covariant angular
decomposition of $I(e^a)$:
\begin{equation}
I(e^a) = \sum_{\ell=0}^{\infty} \Delta_\ell^{-1} I_{A_\ell} e^{A_\ell},
\label{eq:1}
\end{equation}
where $e^{A_\ell}$ is a convenient condensed notation for
$e^{a_1} \dots e^{a_\ell}$, and we have defined
\begin{equation}
\Delta_\ell \equiv \frac{4\pi (-2)^\ell (\ell !)^2}{(2\ell+1)!}.
\end{equation}
Including the factor of $\Delta_\ell$ in~(\ref{eq:1}) has the virtue of
ensuring that the three lowest multipoles $I$, $I_a$ and $I_{ab}$ are
simply the radiation energy density, energy flux, and anisotropic stress,
respectively:
\begin{equation}
\rho^{(\gamma)} = I, \qquad q^{(\gamma)}_a = I_a, \qquad
\pi^{(\gamma)}_{ab} = I_{ab}.
\end{equation}
Propagation equations for the radiation multipoles follow from a multipole
expansion of the Boltzmann equation~\cite{LC-scalcmb,maartens99}.
In linearised form, we have
\begin{eqnarray}
\fl \dot{I}_{A_\ell} + \frac{4}{3}\Theta I_{A_\ell}
+ \uD^b I_{b A_\ell} -\frac{\ell}{(2\ell+1)}\uD_{\langle a_\ell}I_{A_{\ell-1}
\rangle} + \frac{4}{3} \delta_\ell^1 I A_a - \frac{8}{15}
\delta_\ell^2 I \sigma_{a_1 a_2} \nonumber \\
\lo = - \nelec \sigt\left(I_{A_\ell}
- \delta_\ell^0 I - \frac{4}{3} \delta_\ell^1 I v_{a_1}^{(b)}
- \frac{1}{10} \delta_\ell^2 I_{a_1 a_2} \right),
\label{eq:2}
\end{eqnarray}
with $I_{A_\ell}=0$ for $\ell<0$. The projected vector $v_a^{(b)}$ which
appears in the scattering term on the right of~(\ref{eq:2})
is the relative velocity of the baryon
frame, so that $u_a^{(b)} = u_a + v_a^{(b)} + O(\epsilon^2)$ where $u_a^{(b)}$
is the baryon four-velocity. The electron number density in the baryon
frame is denoted $\nelec$, and $\sigt$ is the Thomson cross section.
We have ignored the small effect of polarization
in~(\ref{eq:2}). In~\cite{chall99b} we give a full 1+3 covariant treatment
of CMB polarization. There it is shown that the only effect of polarization on
the intensity multipole equations is to replace $I_{a_1 a_2}$ by
$I_{a_1 a_2} + 6\mathcal{E}_{a_1 a_2}$ in the final term on the
right-hand side of~(\ref{eq:2}), where $\mathcal{E}_{a_1 a_2}$ is a rank-2
PSTF tensor which describes the quadrupole moment of the electric
polarization; see~\cite{hu98} for equivalent results in the total
angular momentum approach.
It is clear from the multipole equations that the complications due to
the anisotropy of Thomson scattering affect only the $\ell=2$ multipole
equation. (This statement is also true of the further complications
due the polarization dependence of Thomson scattering.) This simplicity,
which is preserved by the normal-mode expansion described in
section~\ref{sec_gw}, ensures that the current methodology shares many of the
advantages of the total angular momentum method over traditional Legendre
techniques.

It follows from equation~(\ref{eq:0}) that in linear theory, $\delta_T(e^a)$
has the multipole decomposition~\cite{maartens95}
\begin{equation}
\delta_T(e^a) = \frac{\pi}{I} \sum_{\ell=1}^{\infty}
\Delta_\ell^{-1} I_{A_\ell}e^{A_\ell},
\end{equation}
so that in a statistically isotropic ensemble the power spectrum of temperature
anisotropies, $C_\ell$, where
\begin{equation}
\langle \delta_T(e^a) \delta_T(e^{\prime a}) \rangle =
\sum_{\ell=1}^\infty \frac{2\ell+1}{4\pi} C_\ell P_\ell(\cos \theta),
\end{equation}
with $\cos \theta = -e^a e_a^{\prime}$, can be determined from the covariance
of the radiation multipoles~\cite{gebbie99a}:
\begin{equation}
\left(\frac{\pi}{I}\right)^2 \langle I_{A_\ell}
I^{B_{\ell'}} \rangle = \Delta_\ell C_\ell \delta_{\ell \ell'}
h_{\langle A_\ell \rangle}^{\langle B_\ell \rangle}.
\label{eq:cl}
\end{equation}
Here $h_{\langle A_\ell \rangle}^{\langle B_\ell \rangle} \equiv
h_{\langle a_1}^{\langle b_1} \dots h_{a_\ell \rangle}^{b_\ell \rangle}$,
and large angle brackets denote an ensemble average.

\section{Gravitational waves in an almost-{FRW} universe}
\label{sec_gw}

The 1+3 covariant description of gravitational waves in a cosmological context
has been considered by Hawking~\cite{haw66}, and more recently by
Dunsby, Bassett, \&\ Ellis~\cite{dunsby96a}. Linearised gravitational waves
are described by the transverse degrees of freedom in the electric and
magnetic parts of the Weyl tensor. Similarly, the shear and anisotropic
stress are transverse so that
\begin{equation}
\uD^a E_{ab} = 0,\qquad \uD^a H_{ab} = 0,\qquad \uD^a \sigma_{ab} = 0, \qquad
\uD^a \pi_{ab} = 0
\label{eq:t1}
\end{equation}
at $O(\epsilon)$. Furthermore, the vorticity and all projected vectors that
are generally $O(\epsilon)$ in an almost-FRW universe vanish at linear
order. In particular, in a pure tensor mode the individual matter components
all have the same four-velocity which defines the fundamental velocity $u^a$.

The covariant equations governing the propagation of linearised
gravitational waves can be obtained from the equations of the previous
section. The constraint equations~(\ref{eq:c1}--\ref{eq:c5}) reduce to the
single equation
\begin{equation}
H_{ab} = \curl \sigma_{ab}.
\label{eq:tc}
\end{equation}
which determines the magnetic part of the Weyl tensor from the shear.
The propagation equations~(\ref{eq:p1}--\ref{eq:p7}) reduce to
\begin{eqnarray}
\dot{\rho} & = & - (\rho + p) \Theta \label{eq:tp1} \\
\dot{\Theta} & = & -{\frac{1}{3}}\Theta^2 - {\frac{1}{2}}\kappa
(\rho + 3p) \label{eq:tp2} \\
\dot{\sigma}_{ab} & = & - {\frac{2}{3}} \Theta \sigma_{ab} - E_{ab}
- {\frac{1}{2}}\kappa \pi_{ab} \label{eq:tp3}\\
\dot{E}_{ab} & = & -\Theta E_{ab} + \curl H_{ab} + {\frac{1}{2}}
\kappa \left[ -(\rho + p)\sigma_{ab} + \dot{\pi}_{ab}
+ {\frac{1}{3}} \Theta \pi_{ab}\right] \label{eq:tp4} \\
\dot{H}_{ab} & = & - \Theta H_{ab} - \curl E_{ab} - {\frac{1}{2}}\kappa
\curl \pi_{ab}, \label{eq:tp5}
\end{eqnarray}
Making use of the linearised identities~\cite{maartens98},
valid for an arbitrary first-order PSTF tensor $S_{ab}$,
\begin{eqnarray}
u^b \nabla_b (\curl S_{ab}) = \curl \dot{S}_{ab} - {\frac{1}{3}}
\Theta \curl S_{ab} \label{id:1}\\
u^c \nabla_c (\uD^a S_{ab} ) = \uD^a \dot{S}_{ab} - {\frac{1}{3}}
\Theta \uD^a S_{ab} \label{id:2} \\
\uD^a (\curl S_{ab}) = {\frac{1}{2}} \curl (\uD^a S_{ab}), \label{id:3}
\end{eqnarray}
it is straightforward to verify that the constraints in
equations~(\ref{eq:t1}) and~(\ref{eq:tc}) are consistent with the linearised
propagation equations~(\ref{eq:tp1}--\ref{eq:tp5}).

Inhomogeneous wave equations for the shear and the magnetic part of the
Weyl tensor follow from differentiating equations~(\ref{eq:tp3})
and~(\ref{eq:tp5}) along the flow lines of $u^a$:
\begin{eqnarray}
\fl \ddot{\sigma}_{ab} + \uD^2 \sigma_{ab} + {\frac{5}{3}}
\Theta \dot{\sigma}_{ab} + \left[ {\frac{1}{2}} (4-3\gamma)
\kappa \rho - {\frac{K}{S^2}} \right] \sigma_{ab}
= - \kappa \left[ \dot{\pi}_{ab} + {\frac{2}{3}}\Theta \pi_{ab}
\right] \label{eq:t2} \\
\fl \ddot{H}_{ab} + \uD^2 H_{ab} +  {\frac{7}{3}}\Theta \dot{H}_{ab}
+2 \left[ (2-\gamma)\kappa \rho - {\frac{3K}{S^2}} \right]
H_{ab} = -\kappa \left[\curl \dot{\pi}_{ab} + {\frac{2}{3}}
\Theta \curl \pi_{ab}\right], \label{eq:t3}
\end{eqnarray}
where $\uD^2 \equiv \uD^a \uD_a$, and we have used the linearised identity
\begin{equation}
\curl (\curl S_{ab}) =  \uD^2 S_{ab} + {\frac{3K}{S^2}} S_{ab}
- {\frac{3}{2}} \uD_{\langle a} \uD^c S_{b\rangle c}.
\label{id:5}
\end{equation}
Here, $\gamma$ is defined in terms of the total energy density and pressure:
$p = (\gamma-1)\rho$, $S$ is the scale factor of the universe [defined in
a general almost-FRW model by $\dot{S} = H S$, and $\uD_a S = O(\epsilon)$],
and $6 K/S^2$ is the zero-order intrinsic curvature scalar on comoving
hypersurfaces.

The solutions to the homogeneous wave equations in a $K=0$
universe with constant $\gamma$ were given in~\cite{dunsby96a}.
Solutions for $K\neq 0$ are given in section~\ref{sec_sol} in the radiation
and matter eras. Solutions to the inhomogeneous
equations can be obtained from those solutions by Green's method, but
an accurate treatment of the matter-radiation transition requires a numerical
integration. The damping due to the anisotropic stress terms in the wave
equations was discussed by 
Hawking for the simple viscous model $\pi_{ab} = \lambda \sigma_{ab}$, where
$\lambda$ is the viscosity~\cite{haw66}. During tight-coupling, the radiative
contribution to $\pi_{ab}$ for a general (long wavelength) perturbation is
\begin{equation}
I_{ab} = \frac{16}{27} I\tc (\sigma_{ab} + \uD_{\langle a} v^{(b)}_{b\rangle}
),
\end{equation}
where $\tc = (\nelec \sigt)^{-1}$ is the photon mean free time. Including
polarization changes the factor of $16/27$ to $32/45$ in this expression
for $I_{ab}$, since the electric polarization quadrupole
$\mathcal{E}_{ab} \approx I_{ab}/4$ during tight-coupling.
The tensor $\sigma_{ab}+\uD_{\langle a}v^{(b)}_{b\rangle}$
is the shear in the baryon frame, so the effective (polarization-corrected)
viscosity is $\lambda = 32 I \tc/45$.
Radiative viscous effects, as well as those due to other
(collisionless) species such as neutrinos, lead to a small increase in the
damping of gravitational waves over that due to redshift effects alone.
However, unlike the case of scalar
perturbations, the photon (and neutrino) quadrupole
does not leave a significant imprint on the tensor CMB \emph{temperature}
power spectrum at high multipoles~\cite{durrer98}, since on the scales where
damping due to
anisotropic stresses could be significant, the tensor power spectrum is already
insignificant due to sub-horizon redshifting of the gravity waves.

The wave equation for the
electric part of the Weyl tensor is not closed (even for $\pi_{ab}=0$)
because of the presence of terms involving the shear:
\begin{eqnarray}
\fl \ddot{E}_{ab} + \uD^2 E_{ab} + {\frac{7}{3}}\Theta \dot{E}_{ab}
+ 2\left[(2-\gamma)\kappa\rho - {\frac{3K}{S^2}} \right] E_{ab}
+ {\frac{1}{6}} \kappa \rho \gamma \left[3 {\frac{\dot{\gamma}}{\gamma}}
+ (2-3\gamma)\Theta \right] \sigma_{ab}\nonumber\\
\lo = - {\frac{1}{2}} \kappa \uD^2 \pi_{ab} + \left[
{\frac{2}{3}}\kappa \rho - {\frac{3K}{S^2}}\right]
\kappa \pi_{ab} + {\frac{5}{6}} \Theta \kappa \dot{\pi}_{ab}
+ {\frac{1}{2}} \kappa \ddot{\pi}_{ab}.
\end{eqnarray}
To eliminate the shear requires a further
differentiation~\cite{haw66} (the resulting equation for $\pi_{ab}=0$ was
given in~\cite{dunsby96a}). Although one cannot write down a closed
wave equation for $E_{ab}$, the effective order of the equation is two
since $E_{ab}$ can be determined from the shear via equation~(\ref{eq:tp3}),
and the shear is determined by a wave equation.

\subsection{Mode expansion in tensor harmonics}

In linear perturbation theory it is convenient to expand the $O(\epsilon)$
variables in harmonic modes, since this converts the constraint
equations into algebraic relations and the propagation equations into ordinary
differential equations along the flow lines. For tensor perturbations,
the appropriate harmonic functions can be derived from the tensor
harmonics $Q^{(k)}_{ab}$. The zero-order properties of the tensor harmonics
are discussed in~\ref{appa}, where their explicit PSTF representation
is derived for general $K$.

The electric and magnetic parts of the Weyl tensor, the shear, and the
anisotropic stress can be expanded directly in the electric and magnetic
parity tensor harmonics (see~\ref{appa}):
\begin{eqnarray}
E_{ab} & = &S^{-2}\sum_{k} k^2 (E_k Q^{(k)}_{ab} +
\bar{E}_k \bar{Q}^{(k)}_{ab}) \\
H_{ab} & = &S^{-2}\sum_{k} k^2 (H_k Q^{(k)}_{ab} +
\bar{H}_k \bar{Q}^{(k)}_{ab}) \\
\sigma_{ab} & = & S^{-1}\sum_{k} k (\sigma_k Q^{(k)}_{ab}+
\bar{\sigma}_k \bar{Q}^{(k)}_{ab}) \\
\pi^{(i)}_{ab} & = & \rho^{(i)} \sum_{k} (\pi_k^{(i)} Q^{(k)}_{ab}
+ \bar{\pi}_k^{(i)} \bar{Q}^{(k)}_{ab}),
\end{eqnarray}
where the symbolic $\sum_{k}$ denotes a sum over the harmonic modes.
For closed spatial sections ($K> 0$), the spectrum is discrete with
$\nu \equiv [(k^2+3K)/|K|]^{1/2}$ an integer $\geq 3$. We use an overbar to
distinguish the magnetic parity harmonics from the electric parity, and
a subscript/superscript $k$ to label (implicitly) the distinct eigenfunctions
with degenerate eigenvalue $k$. For the representation in~\ref{appa},
the label $k$ represents the collection $\nu,\ell,m$, where $\ell$ and $m$
describe the orbital angular momentum of the tensor harmonic.
Functions of the eigenvalue alone will
be denoted with $k$ as an argument, for example, $A(k)$, or, equivalently,
$A(\nu)$. The mode coefficients,
such as $E_k$ have $O(\epsilon^2)$ projected gradients by construction.

For the specific representation of the tensor harmonics constructed
in~\ref{appa}, the electric and magnetic parity tensor harmonics are
related by a curl:
\begin{equation}
\curl Q^{(k)}_{ab} = {\frac{k}{S}} \left(1+ {\frac{3K}{k^2}}\right)^{1/2}
\bar{Q}^{(k)}_{ab},
\end{equation}
with an equivalent result for $\curl \bar{Q}^{(k)}_{ab}$. From
equation~(\ref{eq:tc}), we find that $H_k$ is algebraic in $\bar{\sigma}_k$:
\begin{equation}
H_k = \left(1+ {\frac{3K}{k^2}}\right)^{1/2} \bar{\sigma}_k. \label{eq:tc1}
\end{equation}
Note how the presence of curl terms leads to a coupling of the different
polarization states. The constraint equation~(\ref{eq:tc}) [or,
equivalently,~(\ref{eq:tc1})] allows one to eliminate $H_{ab}$
from the discussion, leaving coupled first-order equations for the
electric part of the Weyl tensor and the shear:
\begin{eqnarray}
\fl {\frac{k^2}{S^2}} \left( \dot{E}_k + {\frac{1}{3}}
\Theta E_k \right) - {\frac{k}{S}} \left( {\frac{k^2}{S^2}}
+ {\frac{3K}{S^2}} - {\frac{1}{2}} \gamma \kappa
\rho \right) \sigma_k = - {\frac{1}{6}} (3\gamma - 1)
\kappa \rho \Theta \pi_k + {\frac{1}{2}} \kappa\rho \dot{\pi}_k
\\
\fl {\frac{k}{S}} \left(\dot{\sigma}_k + {\frac{1}{3}}
\Theta \sigma_k \right) + {\frac{k^2}{S^2}}E_k
= - {\frac{1}{2}} \kappa \rho \pi_k.
\label{eq:ph2a}
\end{eqnarray}
Combining these equations gives the harmonic expansion of the wave equation
for the shear [equation~(\ref{eq:t2})]:
\begin{equation}
\fl \ddot{\sigma}_k + \Theta \dot{\sigma}_k + \left[{\frac{k^2}{S^2}}
+ {\frac{2K}{S^2}} -{\frac{1}{3}} (3\gamma-2)\kappa
\rho \right] \sigma_k = \kappa\rho {\frac{S}{k}} \left[
{\frac{1}{3}} (3\gamma -2) \Theta \pi_k - \dot{\pi}_k \right].
\label{eq:ph2}
\end{equation}

For the radiation multipoles, $I_{A_\ell}$, we expand in PSTF tensors
$Q^{(k)}_{A_\ell}$ (and their overbarred counterparts) derived from the
tensor harmonics via
\begin{equation}
Q^{(k)}_{A_\ell} = \left({\frac{S}{k}}\right)^{\ell-2}
\uD_{\langle a_1} \dots \uD_{a_{\ell-2}} Q^{(k)}_{a_{\ell-1} a_\ell \rangle}.
\end{equation}
The factor of $(S/k)^{\ell-2}$ ensures that $\dot{Q}^{(k)}_{A_\ell} =
O(\epsilon)$. For the $\ell$-th multipole,
\begin{equation}
I_{A_\ell} = I \sum_k \beta_\ell^{-1} (I^{(\ell)}_k Q^{(k)}_{A_\ell}
+ \bar{I}^{(\ell)}_k \bar{Q}^{(k)}_{A_\ell}),
\label{eq:ph1}
\end{equation}
where, for later convenience, we have defined
\begin{eqnarray}
\beta_\ell &\equiv& \prod_{n=2}^\ell \kappa_n ,\\
\kappa_\ell &\equiv& [1-(\ell^2-3)K/k^2]^{1/2}, \qquad \ell \geq 2.
\end{eqnarray}
It follows that the photon anisotropic stress $\pi^{(\gamma)}_k =
\kappa_2{}^{-1} I_k^{(2)}$. The $Q^{(k)}_{A_\ell}$ satisfy a number of
differential identities (which are independent of any specific representation
for the tensor harmonics), the most useful of which are
\begin{eqnarray}
\uD^{a_\ell} Q^{(k)}_{A_\ell} = {\frac{(\ell^2-4)}{\ell(2\ell-1)}}
{\frac{k}{S}} \left[ 1- (\ell^2-3){\frac{K}{k^2}}\right]
Q^{(k)}_{A_{\ell-1}} \\
\uD^2 Q^{(k)}_{A_\ell} = {\frac{k^2}{S^2}}\left[1-(\ell-2)(\ell+3)
{\frac{K}{k^2}}\right] Q^{(k)}_{A_\ell}
\end{eqnarray}
for $\ell \geq 2$, and the recursion relation
\begin{equation}
Q^{(k)}_{A_\ell} =  {\frac{k}{S}} \uD_{\langle a_\ell} Q^{(k)}_{A_{\ell-1}
\rangle}
\end{equation}
for $\ell \geq 3$. In a closed universe, the $Q^{(k)}_{A_\ell}$ vanish for
$\ell \geq \nu$, so that only modes with $\nu > \ell$ contribute to
$I_{A_\ell}$. The distinct processes of decomposing $I(e^a)$ into
angular multipoles $I_{A_\ell}$, followed by an expansion of the
multipoles in derivatives of the tensor harmonics, combine to give
a normal mode expansion of the radiation distribution function which
may be regarded as the extension of the Legendre tensor approach,
first introduced by Wilson~\cite{wilson83}, to non-scalar perturbations.
In our opinion, it is one of the virtues of the 1+3 covariant approach that
the physically significant decomposition into angular multipoles $I_{A_\ell}$
is kept distinct from the mathematically convenient harmonic mode
decomposition. Maintaining this distinction allows one to describe the
evolution of the anisotropy in a general cosmological model directly,
without recourse to a harmonic splitting of the perturbations.
Moreover, the present treatment is manifestly gauge-invariant, although we
do not labour this point since gauge-invariance is less of an issue for
tensor modes than for scalars.

Inserting the mode expansion, equation~(\ref{eq:ph1}), into the propagation
equation~(\ref{eq:2}) for the multipoles $I_{A_\ell}$, we find the
mode-expanded multipole equations:
\begin{eqnarray}
\fl\dot{I}^{(\ell)}_k + {\frac{k}{S}}\Biggl[{\frac{(\ell+3)(\ell-1)}{(\ell+1)
(2\ell+1)}}\kappa_{\ell+1}I^{(\ell+1)}_k - {\frac{\ell}{(2\ell+1)}}
\kappa_\ell I^{(\ell-1)}_k\Biggr] - {\frac{8}{15}}{\frac{k}{S}}\delta_\ell^2
\kappa_2 \sigma_k \nonumber \\
\lo = -\nelec\sigt\left(I^{(\ell)}_k-
{\frac{1}{10}}\delta_\ell^2 I^{(2)}_k \right),
\label{eq:ph3}
\end{eqnarray}
with similar equations for the barred variables. Massless neutrinos
satisfy the same multipole equations but with the Thomson scattering
terms omitted. If polarization is included, $I^{(2)}_k$ should be replaced by
$I^{(2)}_k + 6 \mathcal{E}^{(2)}_k$ in the last term on the right-hand side
of~(\ref{eq:ph3}), where $\mathcal{E}^{(2)}_k$ is the coefficient in the
harmonic expansion of the electric polarization quadrupole;
see~\cite{chall99b} for further details.
Equations~(\ref{eq:ph2}) and (\ref{eq:ph3}), plus the
massless neutrino equations, form a closed set of equations that can be
solved to give a complete description of the tensor contribution to the
CMB anisotropy\footnote{When polarization is included, additional
multipole hierarchies must be included for the electric and magnetic
multipoles of the polarization to close the equations~\cite{chall99b}.}.
Analytic solutions for the shear and electric part of the
Weyl tensor are given in the next section, assuming that anisotropic stress
effects are negligible, as well as the integral solution for the
radiation multipoles.

We end this section by noting that in kinetic theory, where the
radiation multipoles may be non-vanishing for all $\ell$, the
familiar decomposition of the perturbations into scalar, vector and tensor
modes is incomplete. A complete set of eigenfunctions for
expanding a rank-$\ell$ PSTF tensor must include the equivalents of the
scalar (rank-0), vector (rank-1), and tensor (rank-$2$) harmonics
up to rank-$\ell$. Since the perturbations in the matter stress-energy
tensor, the kinematic variables, and the Weyl tensor are not affected by the
higher-rank perturbations in linear theory (where perturbations of different
rank evolve independently) the higher-rank perturbations in the distribution
function cannot arise spontaneously. Any power that was initially in
higher-rank perturbations of the photon distribution would thus be damped away
due to Thomson scattering, leaving a negligible effect on the present day
CMB anisotropy.

\section{Analytic solutions for $E_{ab}$, $\sigma_{ab}$ and $I_{A_\ell}$}
\label{sec_sol}

During tight-coupling, the photon anisotropic stress is suppressed by Thomson
scattering, so the only significant contribution to $\pi_{ab}$ is from the
neutrinos. Assuming that this contribution is negligible, it is
straightforward to find the analytic solutions for the Weyl tensor and the
shear during radiation or matter domination. Anisotropic stresses can be
included analytically by constructing the Green's functions from the
solutions of the homogeneous equations given here.

It is convenient to use $x\equiv \surd{|K|} \eta$ as the independent variable,
where $\eta$ is conformal time. During radiation
domination in an open universe, $HS = \surd{|K|}\coth\! x$, so
that~(\ref{eq:ph2}) becomes
\begin{equation}
\frac{\ud^2 \sigma_k}{\ud x^2} + 2\coth\! x \frac{\ud \sigma_k}{\ud x}
+ \left(\nu^2+1-\frac{2}{\sinh^2 x}\right)\sigma_k = 0.
\end{equation}
The solution of this equation which is regular as
$\eta \rightarrow 0$ is
\begin{equation}
\sigma_k \propto \frac{\cos\nu x}{\sinh\!x}-\frac{\sin\nu x \coth\! x}{\nu
\sinh\! x},
\end{equation}
and the irregular solution is
\begin{equation}
\sigma_k \propto \frac{\sin\nu x}{\sinh\!x}+\frac{\cos\nu x \coth\! x}{\nu
\sinh\! x}.
\end{equation}
These solutions for  the shear determine the electric part of the
Weyl tensor from equation~(\ref{eq:ph2a}). 
The regular solution evaluates to
\begin{equation}
E_k \propto \frac{1}{\sqrt{\nu^2+3}\sinh\! x}\left[\frac{\sin\nu x}{\nu}
(\nu^2+1-\coth^2 x) + \cos\nu x \coth\! x \right],
\label{eq:sol1}
\end{equation}
and the irregular solution to
\begin{equation}
E_k \propto \frac{-1}{\sqrt{\nu^2+3}\sinh\! x}\left[\frac{\cos\nu x}{\nu}
(\nu^2+1-\coth^2 x) - \sin\nu x \coth\! x\right]. 
\end{equation}
For $k\eta \ll 1$ (which implies $\nu x \ll 1$ and $x \ll 1$), the mode is
well outside the horizon since $k\eta\ll 1$ implies that
$HS/k \approx 2/(k\eta)$.
When this condition is met the electric part of the Weyl tensor is
approximately constant, with equation~(\ref{eq:sol1}) giving $E_k \propto
2(\nu^2+1)/[3\surd(\nu^2+3)]$.
The solutions for a closed universe can be obtained from
those in an open universe with the substitutions $x\rightarrow ix$ and
$\nu \rightarrow -i\nu$.

During matter domination in an open universe,
$HS=\surd |K|\coth(x/2)$, so that~(\ref{eq:ph2}) becomes
\begin{equation}
\frac{\ud^2 \sigma_k}{\ud x^2} + 2\coth(x/2) \frac{\ud \sigma_k}{\ud x}
+ \left(\nu^2+1-\frac{1}{\sinh^2 (x/2)}\right)\sigma_k = 0.
\end{equation}
The regular solution of this equation is
\begin{equation}
\fl \sigma_k \propto \frac{1}{\sinh^2(x/2)}\left\{\frac{\sin\nu x}{2\nu}
[4\nu^2+1-3\coth^2(x/2)]+3\cos\nu x\coth (x/2) \right\},
\end{equation}
and the irregular solution is
\begin{equation}
\fl\sigma_k\propto\frac{1}{\sinh^2(x/2)}\left\{\frac{\cos\nu x}{2\nu}[4\nu^2+1
-3\coth^2(x/2)]-3\sin\nu x\coth(x/2)\right\}.
\end{equation}
Using equation~(\ref{eq:ph2a}),
the regular solution for the electric part of the
Weyl tensor during matter domination is
\begin{eqnarray}
\fl E_k \propto \frac{1}{\sqrt{\nu^2+3}\sinh^2(x/2)}\Biggl\{
\frac{3\sin\nu x}{2\nu}\coth(x/2)[2\nu^2+1-\coth^2(x/2)] \nonumber \\
-\cos\nu x[2\nu^2+2-3\coth^2(x/2)]\Biggr\},
\end{eqnarray}
and the irregular solution is
\begin{eqnarray}
\fl E_k \propto \frac{1}{\sqrt{\nu^2+3}\sinh^2(x/2)}\Biggl\{
\frac{3\cos\nu x}{2\nu}\coth(x/2)[2\nu^2+1-\coth^2(x/2)] \nonumber \\
+\sin\nu x[2\nu^2+2-3\coth^2(x/2)]\Biggr\}.
\end{eqnarray}
On superhorizon scales, $E_k \propto 2(\nu^2+1)(4\nu^2+1)/[5\surd(\nu^2+3)]$
is approximately constant in the regular mode. In both the matter and radiation
eras, $\sigma_k = O(x)$ in the regular modes on superhorizon scales.
It follows from~(\ref{eq:tc1}) that on superhorizon
scales the Weyl tensor is dominated by the contribution from its electric
part. The solutions given here generalise the $K=0$ solutions derived
in~\cite{dunsby96a}; in the limit $K\rightarrow 0$ we reproduce their
results.

We can provide a formal solution to equation~(\ref{eq:ph3}) for the
intensity multipoles by noting that the
homogeneous equations (obtained by setting $\nelec=0$ and $\sigma_k=0$)
are solved by the functions $\ell(\ell-1)\Phi_\ell^\nu(x)/\sinh^2(x)$
in an open universe, where $x \equiv \surd|K| (\eta_R-\eta)$ with
$\eta_R$ the conformal time at our current position $R$.
Here, $\Phi_\ell^\nu(x)$
are the ultra-spherical Bessel functions~\cite{abbott86}; see~\ref{appa} also.
Defining the optical depth back to cosmic time
$t$, with $t=t_R$ at $R$:
\begin{equation}
\tau(t) \equiv \int_t^{t_R} \nelec\sigt \, \ud t',
\end{equation}
the integral solution for $I^{(\ell)}_k$ in an open universe is
\begin{equation}
\fl I^{(\ell)}_k = \frac{4\ell(\ell-1)}{[(\nu^2+1)(\nu^2+3)]^{1/2}}
\int^{t_R} \ud t'\, e^{-\tau} \left(\frac{k}{S}\sigma_k+\frac{3}{16}
\nelec\sigt \pi^{(\gamma)}_k \right) \frac{\Phi_\ell^\nu(x)}{\sinh^2 x}.
\label{eq:sol2}
\end{equation}
For a closed universe, the hyperbolic functions should be replaced by their
trigonometric counterparts and $\nu^2+n$ by $\nu^2-n$, where $n$ is an
integer. Equation~(\ref{eq:sol2}) can be shown to be equivalent to the tensor
integral solution derived in~\cite{hu98} with the total angular
momentum method. The shear term in~(\ref{eq:sol2}) arises from the
cumulative redshifting effect of shear along the line of sight. The
contribution to the anisotropy at $R$ from the redshift incurred for a given
increment along the null geodesic is weighted by the probability
$e^{-\tau}$ that a  photon subsequently does not scatter before arriving at
$R$. The term involving the radiation anisotropic stress (intensity
quadrupole) in equation~(\ref{eq:sol2}) describes the cumulative
effect of the anisotropic scattering into the beam along the line of sight.
(Inclusion of polarization gives an additional contribution to this term
so that $\pi^{(\gamma)}_k$ should be replaced by $\pi^{(\gamma)}_k +
6 \kappa_2^{-1} \mathcal{E}^{(2)}_k$. This term has only a very small
effect on the intensity multipoles over the range of $\ell$ where the
magnitudes of the $I_{A_\ell}$ are appreciable.)
The geometric factor $\Phi_\ell^\nu(x)/\sinh^2 x$ arises from the projection
of the contraction $Q^{(k)}_{ab}e^e e^b$ of the tensor harmonic with the
projected photon direction at comoving distance $x/\surd|K|$ back along the
line of sight. This follows quite generally from noting that, for any
representation of the tensor harmonics, $Q^{(k)}_{ab}e^a e^b$ at $x$ down
the line of sight is given in terms of the $Q^{(k)}_{A_\ell}e^{A_\ell}$
at the observation point by
\begin{equation}
Q^{(k)}_{ab} e^a e^b |_x = \frac{4\pi}{[(\nu^2+1)(\nu^2+3)]^{1/2}}
\sum_{\ell=2}^\infty \frac{\ell(\ell-1)}{\Delta_\ell \beta_\ell}
\frac{\Phi_\ell^\nu(x)}{\sinh^2 x} Q^{(k)}_{A_\ell} e^{A_\ell}|_R .
\label{eq:sol3}
\end{equation}
This result, which can be regarded as the PSTF extension of the Rayleigh
plane wave result to eigentensors of the Laplacian on a manifold of constant
curvature, can be verified by comparing the derivatives of $Q^{(k)}_{ab}
e^a e^b$ at $x=0$ with those computed from the following recursion relation
for the $Q^{(k)}_{A_\ell}e^{A_\ell}$:
\begin{equation}
e^b \uD_b (Q^{(k)}_{A_\ell}e^{A_\ell}) = {\frac{k}{S}}\left[
Q^{(k)}_{A_{\ell+1}}e^{A_{\ell+1}} - {\frac{(\ell^2-4)}{(4\ell^2-1)}}
\kappa_\ell{}^2 Q^{(k)}_{A_{\ell-1}}e^{A_{\ell-1}} \right]. 
\end{equation}
In a closed universe the sum over $\ell$ in equation~(\ref{eq:sol3})
is restricted to the range $2 \leq \ell \leq \nu-1$.

\section{The {CMB} power spectrum}
\label{sec_pow}

To determine the CMB power spectrum from the $I^{(k)}_\ell$ we require
the $Q^{(k)}_{A_\ell}$ at the observation point $R$. For this purpose, it is
convenient to consider the variation of $Q^{(k)}_{ab}e^e e^b$ down the
line of sight for the chosen representation of the harmonics, and to compare
this result with~(\ref{eq:sol3}) which is independent of the specific
representation. For the representation of the tensor harmonics given
in~\ref{appa}, we find for the electric parity harmonics
\begin{equation}
Q^{(k)}_{ab} e^a e^b|_x = \frac{1}{\nu\sqrt{2(\nu^2+1)}}
\left[\frac{(\ell+2)!}{(\ell-2)!}\right]^{1/2} \frac{\Phi_\ell^\nu(x)}{
\sinh^2 x} \clq_{A_\ell} e^{A_\ell}|_R,
\label{eq:pow1}
\end{equation}
in an open universe, and $\bar{Q}^{(k)}_{ab}e^a e^b|_x =0$ for the magnetic
parity harmonics. Here, $\clq_{A_\ell}$ is a rank-$\ell$ PSTF tensor
encoding the $2\ell+1$ degrees of freedom in the tensor harmonic with
$\ell$ units of orbital angular momentum; see the discussion in~\ref{sec_sum}.
Comparing equations~(\ref{eq:pow1}) and~(\ref{eq:sol3}),
and using the orthogonality of the $e^{\langle A_\ell\rangle}$, we find
\begin{equation}
Q^{(k)}_{A_{\ell'}} |_{R} = \frac{1}{4\pi}
\left[\frac{(\nu^2+3)}{2\nu^2}\right]^{1/2}
\left[\frac{(\ell+2)(\ell+1)}{\ell(\ell-1)}\right]^{1/2}
\Delta_\ell \beta_\ell \clq_{A_\ell}
\delta_{\ell\ell'},
\label{eq:pow2}
\end{equation}
for the electric parity tensor harmonics, and $\bar{Q}^{(k)}_{A_{\ell'}}
|_{R}=0$ for magnetic parity. Although the magnetic parity harmonics do not
contribute to the radiation anisotropy at $R$, they do contribute on other
integral curves of $u^a$. The contribution of the electric parity harmonics
undergoes a compensating change to ensure statistical homogeneity.

Following standard practice we assume that on the scales of cosmological
interest today there is negligible power in the irregular modes given in
section~\ref{sec_sol} during the radiation era. This assumption is justified by
the highly squeezed quantum state predicted at the end of
inflation~\cite{albrecht94,polarski96}.
Since $E_k$ is constant outside the horizon for the regular solution,
it is convenient to parameterise the radiation multipoles in terms
of this constant:
\begin{equation}
I^{(\ell)}_k = \frac{(\nu^2+3)}{(\nu^2+1)}T^{(\ell)}(\nu)
\lim_{k\eta \rightarrow 0} E_k,
\end{equation}
in an open universe. $T^{(\ell)}(\nu)$ is the radiation transfer function,
and the $\nu$-dependent factor is chosen to conform with treatments based
on the metric perturbation variable (e.g.~\cite{hu98}).
The primordial tensor power spectrum $\clp(\nu)$ is defined in terms of the
covariance matrix for the $E_k$ in the limit $k\eta \rightarrow 0$:
\begin{equation}
\langle E_k E_{k'} \rangle = \langle \bar{E}_k \bar{E}_{k'} \rangle
\propto \frac{1}{|K|^{3/2}}\frac{(\nu^2+1)^2}{(\nu^2+3)^2}
\frac{\clp(\nu)}{\nu(\nu^2+1)} \delta_{k k'},
\end{equation}
where $\delta_{kk'}$ is the symbolic delta function appropriate to the sum
over tensor harmonics; see~\ref{appa}. In the $K\rightarrow 0$
limit, $\clp(\nu) \rightarrow k^{n_{\rm T}}$ for a tensor
spectral index $n_{\rm T}$. Making use of equations~(\ref{eq:cl})
and~(\ref{eq:pow2}), we find that the CMB power spectrum
evaluates to
\begin{equation}
C_\ell = \frac{1}{16}\frac{(\ell+2)(\ell+1)}{\ell(\ell-1)}
\int_0^\infty \frac{\nu \ud \nu}{(\nu^2+1)}
\frac{(\nu^2+3)}{2\nu^2} \clp(\nu) [T^{(\ell)}(\nu)]^2
\end{equation}
in an open universe. For closed models the integral over $\nu$ should
be replaced by a discrete sum over integral $\nu \geq \ell+1$.
Note that the CMB power spectrum at multipole $\ell$ depends only
on the transfer functions at the same multipole; a feature not
shared by schemes based on simple Legendre expansions of the distribution
functions (e.g.~\cite{critt93b}).
With these conventions, the contribution to the magnitude (squared) of the
electric part of the Weyl tensor from superhorizon modes is
\begin{equation}
\left({\frac{S}{k}}\right)^4 \frac{\partial}{\partial \ln\! k}
\langle E_{ab} E^{ab} \rangle = \frac{1}{2\pi}
\frac{(\nu^2+4)(\nu^2+1)}{\nu^2 (\nu^2+3)} \clp(\nu).
\end{equation}
For $\nu\rightarrow 0$ in open models, the limit
$\clp(\nu)\rightarrow 0$ is required
to avoid an infrared divergence in the $C_\ell$ and the electric part of the
Weyl tensor. (In bubble models, $\clp(\nu)\rightarrow \nu$ for $\nu\ll
1$~\cite{bucher97}.)

\begin{figure}
\begin{center}
\epsfig{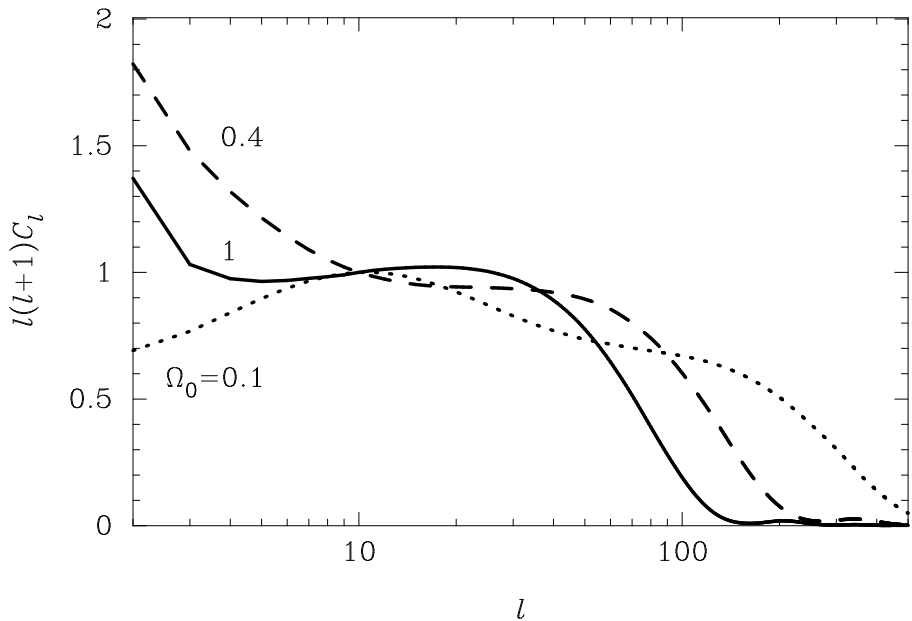}
\end{center}
\caption{Intensity multipoles from gravitational waves in CDM models
with $h=0.7$, $\Omega_{{\rm b}}h^2 = 0.0125$, and no reionisation or
cosmological constant. The $C_\ell$ are shown for two open models,
$\Omega_0=0.1$ and $0.4$, and the flat model.
The primordial power spectrum is $\clp(\nu)=\tanh(\pi\nu/2)$.
The curves are normalised at $C_{10}$.}
\label{fig}
\end{figure}

The $C_\ell$ due to gravitational waves in the CDM model
are plotted in Figure~\ref{fig} for three values of the current-day
matter density: $\Omega_0=0.1$, $0.4$ and $1$. The Hubble constant
$h=0.7$ ($H_0=100h$ km${\rm s}^{-1}$Mp${\rm{c}}^{-1}$), the baryon fraction
$\Omega_{{\rm b}}h^2=0.0125$, and there is no reionisation or cosmological
constant. The primordial power spectrum is the minimal scale-invariant
spectrum $\clp(\nu)=\tanh(\pi\nu/2)$~\cite{bucher97}. For the flat model
($\Omega_0=1$), this $\clp(\nu)\rightarrow 1$ corresponding to
$n_{\rm T}=0$. The power spectra
were calculated by modifying \textsc{CMBFAST}~\cite{seljak96} to solve
the 1+3 covariant equations given in this paper. The results are in excellent
agreement with those given in~\cite{hu97b} which employed the total
angular momentum method.
Notable features in Figure~\ref{fig}, which are discussed
further in~\cite{hu97b}, include the angular scaling brought about
by changes in $\Omega_0$ (variations in the angular diameter distance),
the curvature cutoff in the large scale CMB power spectrum in low
density universes (the curve for $\Omega_0=0.1$),
and the enhancement in power on large scales, compared to the
flat case, for higher values of the density parameter (the curve for
$\Omega_0=0.4$), due to the growth of the shear in the longest wavelength modes
once they enter the horizon.

\section{Conclusion}
\label{sec_conc}

We have developed the 1+3 covariant results necessary to calculate the
temperature anisotropy of the CMB due to gravitational waves in general
almost-FRW models. The
covariant and gauge-invariant approach is well suited to describing
gravitational waves in a cosmological context, since the emphasis is
placed on the physically relevant Weyl tensor and shear, rather than the
perturbation in the metric tensor. In the covariant calculation the
intensity anisotropy is decomposed into irreducible form before
linearisation around an FRW background, so the approach is also well suited
to studying non-linear radiative effects (see~\cite{maartens99}).
For the linear
calculation, the expansion of the radiation intensity in PSTF tensor-valued
multipoles ensures that the subsequent mode expansions of the radiation
variables for scalar, vector and tensor modes naturally generalise the
harmonic expansions of the rank-0, 1 and 2 tensors which describe
inhomogeneity and anisotropy in the geometry.
The same is true for the description of the polarization of the
CMB, as we show elsewhere~\cite{chall99a,chall99b}. The technique should also
allow a straightforward derivation of the multipole hierarchies for massive
particles in difficult cases, such as non-flat models with vector or
tensor perturbations.

\ack
The author acknowledges Queens' College, Cambridge, for support in the form of
a Research Fellowship, and valuable help from Antony Lewis who made the
modifications to the \textsc{CMBFAST} code.

\appendix
\section{The tensor harmonics}
\label{appa}

The tensor harmonics $Q^{(k)}_{ab}$ are zero-order, second-rank PSTF tensors
satisfying
\begin{equation}
\uD^2 Q^{(k)}_{ab} = {\frac{k^2}{S^2}}Q^{(k)}_{ab},
\label{app:1}
\end{equation}
where the equality holds at zero-order. They are constructed to satisfy the
zero-order relations
\begin{equation}
\uD^a Q^{(k)}_{ab} = 0, \qquad \dot{Q}^{(k)}_{ab} = 0,
\label{app:2}
\end{equation}
which are consistent with~(\ref{app:1}) at zero-order. With the definition
$\nu \equiv [(k^2+3K)/|K|]^{1/2}$, the spectrum of regular,
normalisable solutions for
open and flat background models has $\nu \geq 0$, while in closed models
$\nu$ takes integral values $\geq 3$. Supercurvature modes in open models
(see e.g.~\cite{garcia99}) correspond to $0>\nu^2\geq -3$.
For $\nu^2 \gg 1$ the mode is well
inside the curvature radius and $k$ is an effective comoving wavenumber.
We label the tensor harmonics with a lumped superscript $(k)$ which should
be understood to distinguish degenerate solutions of~(\ref{app:1}).

In the text we make use of a specific PSTF representation of the tensor
harmonics. For this representation it is convenient to ``coordinatise''
the spacetime as follows. Starting at some arbitrary point $R$ (which can
conveniently be taken to be our current location), we construct spatial
geodesics with tangent vectors $e^a(\chi)$, the generators of which at
$R$ are the set of projected unit vectors $\{e^a|_R\}$ at $R$:
\begin{equation}
e^a\uD_a e^b =0, \qquad u^a e_a = 0, \qquad \frac{\ud x^a}{\ud \chi} =
-S e^a,
\label{app:3}
\end{equation}
with $\chi \geq 0$ (the equality holding at $R$).
In the absence of vorticity the curves $x^a(\chi)$ lie in the flow-orthogonal
hypersurface through $R$. A projected vector field $e^a$ and a
scalar field $\chi$ can then be constructed in the neighbourhood of
$R$ from their restriction on the hypersurface defined by $\{e^a|_R,\chi\}$ via
\begin{equation}
\dot{e}^{\langle a \rangle} = 0, \qquad \dot{\chi} = 0.
\label{app:4}
\end{equation}
The fields $e^a$ and $\chi$ satisfy equations~(\ref{app:3}) at
zero-order in an almost-FRW universe. To the same order, the restriction of
$e^a$ to the past lightcone through $R$ is the projected direction of
a photon that passes through $R$, and $\chi$ is the zero-order
conformal lookback time.

\subsection{Electric parity harmonics}

For the electric parity tensor harmonics we start with the most general
electric parity, rank-two PSTF tensor~\cite{thorne80} which is constant
along the flow lines of $u^a$:
\begin{eqnarray}
\fl Q^{(k)}_{ab} = T_1(x) \left(e_a e_b \clq_{C_\ell} e^{C_\ell} +
{\frac{1}{2}} \clh_{ab} \clq_{C_\ell} e^{C_\ell} \right) \nonumber \\
+ T_2(x) e_{(a} \clh_{b)}^{c_\ell}
\clq_{C_\ell} e^{C_{\ell-1}} + T_3(x) [\clq_{ab C_{\ell-2}}
e^{C_{\ell-2}}]^{\TT},
\qquad \ell\geq 2. \label{app:5}
\end{eqnarray}
Here, $\clq_{A_\ell}$ is a rank-$\ell$ PSTF tensor field satisfying the
zero-order equations
\begin{equation}
e^b \uD_b \clq_{A_\ell} =0, \qquad \dot{\clq}_{A_\ell} = 0,
\label{app:6}
\end{equation}
which determine the tensor field $\clq_{A_l}$ from its value at
the point $R$, and $T_i(x)$, $i=1,2,3$, are scalar functions of
$x \equiv \surd|K|\chi$. The (screen) projection tensor $\clh_{ab}$ is
\begin{equation}
\clh_{ab} \equiv h_{ab} + e_a e_b,
\label{app:7}
\end{equation}
and the notation $[A_{ab}]^\TT$ denotes the transverse traceless
part of the second-rank tensor $A_{ab}$:
\begin{equation}
[A_{ab}]^\TT \equiv \clh_a^{c_1} \clh_b^{c_2} A_{c_1 c_2} -
{\frac{1}{2}} \clh_{ab} \clh^{c_1 c_2} A_{c_1 c_2}.
\label{app:8}
\end{equation}
Demanding that $\uD^a Q^{(k)}_{ab} = 0$ determines $T_2$ and $T_3$ in
terms of $T_1$. In an open universe we find
\begin{eqnarray}
T_2(x) &=& \frac{-2}{(\ell+1)\sinh^2\!x} \frac{\ud}{\ud x}
[\sinh^3\!x T_1(x)],\label{app:9} \\
T_3(x) &=& \frac{-\ell}{(\ell+2)} T_1(x) -
\frac{1}{(\ell+2) \sinh^2\!x} \frac{\ud}{\ud x}[\sinh^3\!x T_2(x)].
\label{app:10}
\end{eqnarray}
Equation~(\ref{app:1}) gives three further equations,
which are satisfied by virtue of equations~(\ref{app:9}) and~(\ref{app:10})
provided that
\begin{equation}
\frac{\ud^2}{\ud x^2}T_1(x) + 6 \coth\!x \frac{\ud}{\ud x}T_1(x)
+ \left[\nu^2+9 - \frac{(\ell^2+\ell-6)}{\sinh^2\!x} \right]
T_1(x)=0.\label{app:11}
\end{equation}
The regular, normalisable solution of this equation is
\begin{equation}
T_1(x) = N(\nu) \frac{\Phi_\ell^\nu(x)}{\sinh^2\!x},
\label{app:12}
\end{equation}
with $\nu \geq 0$ (subcurvature). The $\Phi^\nu_\ell(x)$ are the
ultra-spherical Bessel functions~\cite{abbott86} defined recursively from
\begin{equation}
\frac{\ud}{\ud x} \Phi_\ell^\nu(x) = \ell \coth\! x \Phi_\ell^\nu(x)
-[\nu^2 + (\ell+1)^2]^{1/2} \Phi^\nu_{\ell+1}(x),
\label{app:13}
\end{equation}
where
\begin{equation}
\Phi_0^\nu(x) = \frac{\sin\!\nu x}{\nu\sinh\! x}.
\label{app:14}
\end{equation}
We choose the normalisation constant $N(\nu)$ so that
\begin{equation}
\int \ud \Omega_{e^a|_R} \ud x\sinh^2\!x
Q^{(k)}_{a_1 a_2} Q^{(k')a_1 a_2}
= \frac{\pi}{2}\delta_{\ell\ell'} \Delta_\ell \clq_{A_l} \clq^{\prime A_l}
\nu^{-2} \delta(\nu - \nu'),
\label{app:15}
\end{equation}
which gives
\begin{equation}
N(\nu) = \frac{1}{\nu\sqrt{2(\nu^2+1)}}
\left[\frac{(\ell+2)!}{(\ell-2)!}\right]^{1/2}.
\label{app:16}
\end{equation}
For closed universes, the hyperbolic functions should be replaced by their
trigonometric counterparts in equations~(\ref{app:9}--\ref{app:16}),
$\nu^2+n$ should be replaced by $\nu^2-n$ with $n$ an integer, and
$\delta(\nu - \nu')$ should be replaced by $\delta_{\nu \nu'}$.
For closed models, the regular, normalisable modes have $\nu$ an integer
$\geq 3$, restricted to $\nu > \ell$.

\subsection{Magnetic parity harmonics}

For the magnetic parity tensor harmonics, which we denote with an overbar,
we start with the general magnetic parity, rank-two PSTF
tensor~\cite{thorne80}:
\begin{equation}
\bar{Q}^{(k)}_{ab} = \bar{T}_1(x) [e_{c_\ell} {\epsilon^{c_\ell
c_{\ell-1}}}_{(a}\clq_{b)C_{\ell-1}} e^{C_{\ell-2}}]^{\TT}
+ \bar{T}_2(x) e_d e_{(a} {\epsilon_{b)}}^{d c_\ell} \clq_{C_\ell}
e^{C_{\ell-1}},
\label{app:17}
\end{equation}
with $\ell\geq 2$. Equations~(\ref{app:1}) and~(\ref{app:2}) give
\begin{equation}
\bar{T}_1(x) = \frac{-1}{(\ell+2)\sinh^2\!x} \frac{\ud}{\ud x}
[\sinh^3\!x \bar{T}_2(x)],
\label{app:18}
\end{equation}
and
\begin{equation}
\frac{\ud^2}{\ud x^2}\bar{T}_2(x) + 4 \coth\!x \frac{\ud}{\ud x}\bar{T}_2(x)
+ \left[\nu^2+4-\frac{(\ell^2+\ell-2)}{\sinh^2\!x} \right]
\bar{T}_2(x)=0.
\label{app:19}
\end{equation}
The regular, normalisable solution of~(\ref{app:19}) is
\begin{equation}
\bar{T}_2(x) = \bar{N}(\nu) \frac{\Phi_\ell^\nu(x)}{\sinh\!x},
\label{app:20}
\end{equation}
where $\nu \geq 0$. With the same normalisation as the electric parity
harmonics, we find
\begin{equation}
\bar{N}(\nu) = \frac{1}{(\ell+1)} \left[\frac{2}{(\nu^2+1)}\right]^{1/2}
\left[\frac{(\ell+2)!}{(\ell-2)!}\right]^{1/2},
\label{app:21}
\end{equation}
so that $\bar{N}(\nu)/N(\nu)=2\nu/(\ell+1)$. In a closed universe we make the
same replacements as for the electric parity harmonics.

As a consequence of the first-order identity
\begin{equation}
\uD^2 (\curl S_{ab}) = \curl (\uD^2 S_{ab}),
\label{app:22}
\end{equation}
where $S_{ab}$ is an $O(\epsilon)$ PSTF tensor, it follows that
if $Q^{(k)}_{ab}$ solves equation~(\ref{app:1})
then so does $\curl Q^{(k)}_{ab}$.
Furthermore, by virtue of equations~(\ref{id:1}) and~(\ref{id:3}),
$\curl Q^{(k)}_{ab}$ will also satisfy~(\ref{app:2}) if $Q^{(k)}_{ab}$
does. Since the curl operation is parity reversing, we expect the
electric and magnetic parity tensor harmonics for given $\ell$ to be
related by a curl. With the normalisation conventions
adopted here, it is straightforward to show that
\begin{eqnarray}
\curl Q^{(k)}_{ab} &=& {\frac{k}{S}} \left(1+{\frac{3K}{k^2}}\right)^{1/2}
\bar{Q}^{(k)}_{ab},
\label{app:23}\\
\curl\bar{Q}^{(k)}_{ab}&=&{\frac{k}{S}}\left(1+{\frac{3K}{k^2}}\right)^{1/2}
Q^{(k)}_{ab}, \label{app:24}
\end{eqnarray}
which are consistent with the first-order identity~(\ref{id:5}).

\subsection{Sum over harmonic modes}
\label{sec_sum}

A normalisable, first-order, rank-2 PSTF tensor field, such as the electric
part of the Weyl tensor, can now be expanded in terms of the electric and
magnetic parity tensor harmonics. Since the $\clq_{A_\ell}$ have only
$2\ell+1$ degrees of freedom, it is convenient to introduce a set of
$2\ell+1$ orthogonal basis PSTF tensors $\{\clq_{A_\ell}^{(\ell m)}\}$, with
$m = -\ell, \dots ,\ell$. We choose the basis tensors so that
\begin{equation}
\clq_{A_\ell}^{(\ell m)} \clq^{(\ell m')A_\ell} =
\Delta_\ell^{-1} \delta_{m m'},
\label{app:26}
\end{equation}
which implies
\begin{equation}
\sum_m \clq_{A_\ell}^{(\ell m)} \clq^{(\ell m)B_\ell} = \Delta_\ell^{-1}
h_{\langle A_\ell \rangle}^{\langle B_\ell \rangle}.
\label{app:27}
\end{equation}
The tensor harmonic expansion of $E_{ab}$ can then be written as
\begin{equation}
E_{ab}=\int_0^\infty |K|^{3/2} \nu^2 \ud \nu\, \sum_{\ell=2}^{\infty} 
\sum_{m=-\ell}^{\ell} (E_{\nu \ell m} Q^{(k)}_{ab} + \bar{E}_{\nu \ell m}
\bar{Q}^{(k)}_{ab}),
\label{app:28}
\end{equation}
in an open universe. [Equation~(\ref{app:28}) can easily be generalised to
include supercurvature modes, if these are present in the initial conditions.]
In the text we
use the abbreviated notation $E_k$ for $E_{\nu \ell m}$, as we have done with
the labelling of the tensor harmonics themselves. The summation and integral
in~(\ref{app:28}) defines the symbolic summation over harmonic modes,
$\sum_k$, employed in the text. In a closed universe, equation~(\ref{app:28})
should be replaced by
\begin{equation}
E_{ab}=\sum_{\nu =3}^\infty |K|^{3/2} \nu^2 \sum_{\ell =2}^{\nu-1}
\sum_{m=-\ell}^{\ell}(E_{\nu \ell m} Q^{(k)}_{ab} + \bar{E}_{\nu \ell m}
\bar{Q}^{(k)}_{ab}). 
\label{app:29}
\end{equation}
Under the assumptions of statistical homogeneity and isotropy, the covariance
structure of the $E_k$ and $\bar{E}_k$ is restricted to the form
\begin{eqnarray}
\langle E_{k} E_{k'} \rangle &=& E^2(\nu) \delta_{kk'} \nonumber \\
\langle \bar{E}_{k} \bar{E}_{k'} \rangle &=& E^2(\nu) \delta_{kk'} \nonumber \\
\langle E_{k} \bar{E}_{k'} \rangle &=& 0, \label{app:30}
\end{eqnarray}
where the symbolic $\delta_{kk'}$ represents $\delta_{\ell\ell'}
\delta_{m m'}|K|^{-3/2}\nu^{-2} \delta(\nu-\nu')$ in an open universe.
In the closed case $\delta(\nu-\nu')$ should be replaced by
$\delta_{\nu\nu'}$ in this expression.
%
%

%

\end{document}